\documentclass[12pt]{article}

\begin{document}
\begin{center}
\Large Applying Popper's Probability

\vspace{1.0cm}

\normalsize
{\it by Alan B. Whiting \\
University of Birmingham}
\end{center}
\begin{abstract}
Professor Sir Karl Popper (1902-1994) was one of the most influential
philosophers of science of the twentieth century, best known for his
doctrine of falsifiability.  His axiomatic
formulation of probability, however, is unknown to current scientists,
though it is championed by
several current philosophers of science as superior to the familiar version.
Applying his system to problems identified by himself and his
supporters, it is shown that it does not have some features he
intended and does not solve the problems they have identified. 
\end{abstract}

\vspace{1.0cm}

\section{Probability and the philosophers}

Professor Sir Karl Popper (1902-1994) is known to scientists as the author
of the `doctrine of falsifiability,' in which a statement is only admitted
to be scientific if it can, in principle, be falsified.  Although not
strictly the originator of the idea, he can be credited with emphasizing
it and it is a useful test for pseudoscientific statements.
His clearest statement of this is
from the {\em Postscript} to {\em The Logic of Scientific Discovery}:

\begin{quote}
\ldots we adopt, as our criterion of demarcation, the {\em criterion
of falsifiability, i.\ e.\ } of an (at least) unilateral or asymmetrical
or {\em one-sided} decidability.  According to this criterion, statements,
or systems of statements, convey information about the empirical world
only if they are capable of clashing with experience; or more precisely,
only if they can be {\em systematically tested}, that is to say, if
they can be subjected (in accordance with a `methodological decision')
to tests which {\em might} result in their refutation.  \footnote{The work
is available in at least two editions, either of which may be conveniently
available but which differ slightly in pagination.  In what follows
I will cite the section; then the page numbers in [1]; then those in
[2].} \S *i, pp.\ 313-4, p.\ 315.  [Here I use
his italics, as I will henceforth.]
\end{quote}

There is a great deal of
other material in this, his most famous work, however; a previous
investigation [3] has shown that it demonstrates
Popper's understanding of science
and especially mathematics to be often inadequate or erroneous.  
Here I will examine one part of the work that has been singled out as useful.

A significant part of {\em The Logic of Scientific Discovery} is given
over to an axiomatic formulation of probability.
It is unknown among practicing
scientists, but a number of philosophers cite it with approval,
considering it superior to the familiar version. 

Fitelson, Haj\'{e}k and Hall [4] consider Popper's formulation
`the most general and elegant' of proposed axiomatizations of two-place
probability (that is, formulations in which conditional probability
is taken as elementary, and absolute or one-place probability is a
derived form).  McGee [5] agrees that `Popper's axioms constitute a
true generalization of ordinary probability theory.'
Roeper and Leblanc [6] note that autonomous formulations of probability,
that is those constructed independent of semantic constraints,
`pioneered by Popper, are preferred by us.' (p. 5).  A clue as to what
the generalization consists of is given by the same authors when they
state
\begin{quote}
It is advantageous to turn Carnap's partial functions into total ones,
this by stipulating that P(A,B)=1 when B is logically false, an
option discussed by Carnap in (1950) and widely adopted in studies
of the subject. (p. 11)\footnote{Notation among the references is
almost uniform, but not quite.  Letters, upper or lower case, are
events or hypotheses, things that might be true or not. P(A,B), with
the letter `P' upper or lower case, is the probability of event A
given that B is true.  The probability of
a combination of events, that is the probability
of both A and B happening given C, is P(AB,C).
Converting to the form more usual for scientists,
$p(a \mid b)$, is straightforward.  In my own expositions I will follow
that used by Popper, $p(a,b)$, to make comparison easier.}
\end{quote}

That is, Roeper and Leblanc desire a formulation of probability in which
{\em any} statement (A) is true (probability unity), 
if a false statement (zero probability, B) is given as a condition.

Perhaps the clearest statement championing Popper, however, is
given by Alan Haj\'{e}k [7].  His overall thesis is that conditional
probability is basic and absolute probability a derived notion,
something Popper provides.  He also identifies some
problems with the conventional formulation that Popper's version
is claimed to solve.  These are centered
on situations with zero probability.

Using the normal formulation, the probability of two things {\it A}
and {\it B} both occurring is
\begin{equation}
p(AB) = p(A,B) p(B)
\end{equation}
which is quickly rearranged to form the conditional probability,
that of {\it A} given {\it B}:
\begin{equation}
p(A,B) = \frac{p(AB)}{p(B)}
\end{equation}

Haj\'{e}k finds this inadequate in two types of problems.  First,
the sphere:
\begin{quote}
A point is chosen at random from the surface of the earth (thought of
as a perfect sphere); what is the probability that it lies in
the Western hemisphere given that it lies on the equator? 1/2, surely.
Yet the probability of the condition is 0, since a uniform probability
measure over a sphere must award probabilities to regions in proportion
to their area, and the equator has area 0.  The ratio analysis thus
cannot deliver the intuitively correct answer.  (pp. 111-112)
\end{quote}
Next, there is the situation of all possible sequences of infinite
tosses of a coin.
\begin{quote}
Any particular sequence has probability zero (assuming that the trials
are independent and identically distributed with intermediate probability 
for heads).  Yet surely various corresponding conditional probabilities
are defined---e.\ g.\ , the probability that a fair coin lands heads on
every toss, given that it lands heads on tosses 3, 4, 5, \dots, is 1/4.
(p. 112)
\end{quote}

(In other places Haj\'{e}k makes more extensive attacks on conventional
probability, but this is where they are most clearly connected with 
Popper.)  Here Haj\'{e}k seeks a system that gives a defined probability
conditional upon an event with zero probability, something he finds in
Popper.

So Popper's formulation of probability is
currently considered a viable and even superior version among at least
a group of philosophers. 

\section{Popper's motivation}

It is instructive to gain a flavor of Popper's reasoning by examining
his motivation for developing a new formulation of probability.

In \S 80 of {\em the Logic of Scientific
Discovery} Popper is concerned with constructing the probability
of a hypothesis from a series of events, or rather with refuting
Richenbach's attempt to do so.  He notes that, if one makes a ratio
of events confirming the hypothesis to total events, one finds
a `probability' of 1/2 for a hypothesis that is refuted by half
the events (p. 257, p. 255).  After trying several variations on
this theme, he concludes, `This seems to me to exhaust the possibilities
of basing the concept of the probability of a hypothesis on that
of the frequency of true statements (or the frequency of false ones),
and thereby on the frequency theory of the probability of events
(p. 260, p. 258).'

In the {\em Postscript}, \S *vii, Popper modifies this conclusion,
stating
\begin{quote}
We may to this end interpret the universal statement {\it a}
as entailing an infinite product of singular statements, each endowed
with a probability which of course must be less than unity. 
(p. 364, p. 376)
\end{quote}
Popper does not explicitly connect these `singular statements' to the
more familiar language of a theory and its predictions.  However, from
this and other sections it seems clear that he is asserting that
the probability of a theory is the product of the probabilities
of all its predictions.  He realises that this implies that all the
`singular statements' must be independent, if considered as probabilities
of events, but justifies his formulation
with the opaque assertion that any other
situation is `non-logical' (pp. 367-8, p. 379).  Popper's conclusion, 
in an infinite
universe, is that {\em all} theories are of zero probability 
(p. 364, p. 376)\footnote{It has been pointed out to the author that
an infinite product of factors all less than one is not necessarily
zero.  If $a_n = (1-1/n^2)$, for instance, and one begins at $n=2$,
the product converges to 1/2.  This is
less important in the present context than the fact that Popper's
whole approach is erroneous.}.

A separate route leading to the same conclusion starts from a
development in Jeffreys' {\em Theory of Probability} ([9], pp. 38-9).
Jeffreys starts with the hypothesis $q$, previous information $H$
(which is not, as Popper notes, vital to the argument) and some
experimental fact $p_1$.  Writing $P(q p_1,H)$ two ways and
rearranging, 
\begin{equation}
P(q,p_1 H) = \frac{P(q,H)P(p_1, qH)}{P(p_1, H)}
\end{equation}
Now if $p_1$ is a consequence of $q$, $P(p_1, qH) = 1$ so
\begin{equation}
P(q,p_1 H) = \frac{P(q,H)}{P(p_1, H)}.
\end{equation}
If $p_2, p_3, \ldots, p_n$ are further consequences of $q$ which are each
in turn found to be true,
\begin{equation}
P(q,p_1 p_2 \ldots H) = \frac{P(q,H)}
{P(p_1,H)P(p_2,p_1H)\cdots P(p_n, p_1 p_2 \cdots p_{n-1}H)}
\label{Jeffreys}
\end{equation}
in which we are dividing $P(q,H)$ by a growing product of probabilities.
Jeffreys notes three possibilities for $P(q,p_1 p_2 \ldots H)$: (1)
it can grow without limit, as it will
if the (unbounded in number) series
of $P(p_n, \cdots)$ has a significant population less than unity;
(2) $P(q,H)$ may be identically zero, in which case $P(q,p_1 p_2 \ldots H)$
will also be identically zero; or (3) the $P(p_n, \cdots)$ become
arbitrarily close to unity.  The first option is ruled out by
the definition of probability.  Jeffreys choses option (3), interpreting
it as a growing confidence in the predictions of a theory based on
a growing number of successful predictions.

In \S *vii (pp. 370-1, pp. 383-4), Popper asserts that (3) leads to a paradox.
He adduces two hypotheses (call them $q_1$ and $q_2$), each of which
predict $p_1,p_2, \ldots, p_{n-1}$.  But while $q_1$ predicts $p_n$,
$q_2$ predicts its contradiction, $\bar{p_n}$.  Then Jeffrys'
formulae predict {\em both} $p_n$ and $\bar{p_n}$ with near-unity
probability.  Hence the only choice is (2), all theories have zero
probability. 

Of course there is no paradox in $P(p_n,p_1 \cdots H q_1) \simeq 1$
at the same time as $P(\bar{p_n},p_1 \cdots H q_2) \simeq 1$; the
probabilities have different conditions.  (It is interesting that, in
a section devoted to the probability of hypotheses, Popper ignores them.)
In fact this is the classical
decisive experiment, which actually happens much more rarely than the
tidy-minded would like. 

There are things about Jeffrys' calculation to make one uneasy.  It
is rather circular, for instance, to stipulate that the $p_n$ all
actually happen (so that $P(p_n, \cdots)=1$), 
then conclude that their probabilites are high.  But one cannot use
it to conclude that the probabilities of all theories vanish.

In these three sections, then, we have examples of Popper using
the conventional theory of probability to calculate
the probability of a theory, based on its predictions and events,
and getting it wrong.  It is worth noting that he has seen it done
properly in at least one instance, and accepts it as correct.  In
\S *ix (pp. 407-8, p. 425) he asserts that the probability of a coin
landing heads is 1/2, given that it has landed heads in
500,000 $\pm$ 1,350 tosses out of a million previously.  The
tacit hypothesis `this is a fair coin' has been given a high probability
based on the probability of its predicted events, and {\em not} by forming
a product of a million factors of 1/2.
  
As a consequence of Popper's reasoning, he asserts,
`\ldots there is a need for a probability
calculus in which we may operate with second arguments of
zero absolute probability' (\S*vi, p. 330, p. 335).  Note that his motivation,
based on errors in applying the conventional system of probability, is
not the same as that of Haj\'{e}k's problems [7] or the `total functions'
of Fitelson, Haj\'{e}k and Hall [4], though they lead to the same
requirement. 

We will add one further problem Popper identifies with conventional
probabily as applied to the confirmation of hypotheses.  
It is not directly related to the zero-probability matter, but forms
another aspect of Popper's attack and we may apply his
system to it.
In \S*ix
(pp. 390-1, pp. 406-7) he presents the situation of a standard six-sided
die.  Hypothesis $x$ is that it rolls a six, and $\bar{x}$ that it
rolls some other number.  Given a fair die (tacitly assumed by
Popper) and no other information, $P(x)=1/6$, $P(\bar{x})=5/6$.
Then given the information $z$ that the die roll was even, we
have $P(x,z)=1/3$, $P(\bar{x},z)$=2/3.  The added information has
increased the probability of $x$ and decreased that of $\bar{x}$, 
but still $P(x,z)<P(\bar{x},z)$.  Popper finds this `clearly
self-contradictory' if probability is to be used to judge the
corroboration of a theory.  Popper is requiring that {\em any}
evidence that supports a theory makes it more likely  than its
contradiction.  Let us call this the requirement for absolute support.

\section{Developing Popper's formulation}

It would be very useful to have a demonstration of exactly how
Popper's system of probability is used in a given instance, and
especially one showing its advantages over the conventional one.
Unfortunately, in the references at hand
the authors do not set out any details about how Popper's
formulation solves the problems they identify.  Nor does Popper
show, in any specific problem, where his formulation solves or
even addresses the flaws he identifies with conventional
probability.  We must work that out
from the axioms themselves, an approach that also permits us to see what the
formulation says and does not say.  After some development, we
will then apply the system to the several problems.

Popper sets out his axioms in slightly different ways in several places
in [1],[2], but the development here (as well as the exposition in
[7]) generally follows \S *iv, pp. 332ff, pp. 336ff; \S *v, pp. 349-353,
pp. 356-361.  Although everything in the formulation can be 
immediately interpreted
in terms of conditional probability, we will refrain from any such
interpretation until after a number of results have been obtained
formally.

We begin with {\it S}, a collection (one would say {\em set}, but
Popper is mistaken on important parts of set theory [3] so we avoid the term) 
of otherwise undefined objects {\em a, b, \ldots}.
There is a unary operation of complementation, with $\bar{a}$
also being in {\it S}, and two binary operations: conjunction,
with $ab$ also a member of {\it S}, and $p(a,b)$ being a real number.

The axioms, with {\it a, b, c, d, \ldots} representing any member of 
{\it S}, are:

\noindent {\bf A1}.  There are elements {\it a, b, c} and {\it d} in {\it S}
such that $p(a,b) \neq p(c,d)$.

\noindent {\bf A2}.  If $p(a,c)=p(b,c)$ for every {\it c} in {\it S},
then $p(d,a) = p(d,b)$ for every {\it d} in {\it S}.

\noindent {\bf A3}.  $p(a,a) = p(b,b)$

\noindent {\bf B1}. $p(ab,c) \leq p(a,c)$

\noindent {\bf B2}.  $p(ab,c) = p(a,bc)p(b,c)$

\noindent {\bf C}i.  $p(a,b) + p(\bar{a}, b) = p(b,b)$, if there is
some {\it c} such that $p(c,b) \neq p(b,b)$.

\noindent {\bf C}ii.  $p(\bar{a}, b) = p(a,a)$ if there is no such {\it c}.

Note in particular the exclusive nature of Ci and Cii. 

First, we set bounds on $p(a,a)$, invoking A3, B1 and B2:

\begin{eqnarray*}
p(a,a) &=& p(b,b) = k \\
p((aa)a,a) &\leq& p(aa,a) \\
p(aa,a) &\leq& p(a,a) = k \\
p((aa)a,a) &=& p(aa,aa)p(a,a) = k^2 \\
k^2 &\leq& k \\
0 &\leq& k \leq 1
\end{eqnarray*}

Next, some bounds on the general $p(a,b)$ and a useful result on
arguments in both positions.  Invoking B1 and B2:

\begin{eqnarray*}
p(ab,ab) &\leq& p(a,ab) \\
k &\leq&p(a,ab) \\
p(aa,b) &=& p(a,ab) p(a,b) \\
p(a,b) &\geq& p(a,ab) p(a,b) \\
1 &\geq& p(a,ab), p(a,b) \neq 0 
\end{eqnarray*}
Now using B1 and B2 again,
\begin{eqnarray*}
p(ab,c) &=& p(a,bc)p(b,c) \\
p(b,c) &\geq& p(a,bc)p(b,c) \\
1 &\geq& p(a,bc), p(b,c) \neq 0
\end{eqnarray*}

Popper derives $k=1$ using Ci. To save time and because we want to avoid
depending on a particular branch of C, we shall simply assume
$p(a,a)=1$, which of course must be true if we are to interpret these
axioms as a formulation of probability.  With the above development,
that means
\begin{equation}
p(a,ab) = 1
\label{eq:repeat}
\end{equation}

It is convenient to have a version of eq. \ref{eq:repeat} with the
conditional reversed (note we have not yet shown that $p(\cdot,ab)
=p(\cdot,ba)$).
Starting with B2,
\begin{equation}
p(ab,c) = p(a,bc) p(b,c) \nonumber
\end{equation}
but since $p(a,bc)$ is bounded by unity,
\begin{equation}
p(ab,c) \leq p(b,c) 
\end{equation}
and so
\begin{equation}
p(ab,ab) \leq p(b,ab) \nonumber
\end{equation}
and, with $k=1$,
\begin{equation}
p(b,ab)=1.
\label{eq:repeat2}
\end{equation}

Now using Ci (and employing $b$ as a general member of $S$ to avoid potential
confusion in what follows),
\begin{eqnarray*}
p(b,b) + p(\bar{b}, b) &=& p(b,b) \\
p(\bar{b},b) &=& 0
\end{eqnarray*}
Note that in using Ci we have made an assumption about the behaviour of
$b$; the following shows it isn't necessarily true.

We again write Ci,
\begin{equation}
p(a, a\bar{a}) + p(\bar{a}, a\bar{a}) = p(a\bar{a},a\bar{a})
\end{equation}
which, in light of equations \ref{eq:repeat} and \ref{eq:repeat2}, 
gives $2=1$; so Ci cannot
hold and we are forced into Cii:
\begin{equation}
p(b, a\bar{a}) = 1
\label{eq:abara}
\end{equation}
for any $b$ in $S$. 

Note that the elements of $S$ have been divided into two classes by
Popper: those for which Ci is true and  $p(\bar{a},a) = 0$,
and those for which Cii is true and $p(\bar{b},b) = 1$.  All
elements of $S$ of the form $a\bar{a}$ are in the latter class.

At this point we may impose the obvious interpretation of Popper's 
system as a formulation of probability.  The
elements of $S$ are events, things that happen or not; they may be
composite, $ab$ meaning that both $a$ and $b$ happen; and $p(a,b)$ is
the probability that $a$ occurs, given that $b$ does.  A Ci event and
its complement exhaust all possibilities, and either
$b$ or $\bar{b}$ is true.  Cii events, however, do unconventional things,
which we shall examine in a moment.

\section{Applying Popper}

As a warm-up, we will apply Popper's system to his requirement of 
absolute support.  That is the requirement that any event 
or information making an hypothesis more
probable than before must make it more probable than its contradiction.
In our notation, we require
\begin{equation}
p(a,bc)>p(a,c) \Rightarrow p(a,bc)>p(\bar{a}, bc)
\end{equation}
(note that the inequalities preclude $bc$ or $c$ being Cii events).  This
means
\begin{equation}
p(a,c) \geq p(\bar{a},bc)
\label{require}
\end{equation}
since if $p(a,c) < p(\bar{a},bc)$ there are possible values for
$p(a,bc)$ that would allow
\begin{equation}
p(a,c) < p(a,bc) < p(\bar{a},bc)
\end{equation}
and absolute support is not satisfied.
Applying B2 to Eq.~\ref{require} and rearranging,
\begin{equation}
p(b,c)p(a,c) \geq p(\bar{a}b,c)
\label{middle}
\end{equation}
Next we write a version of Ci:
\begin{equation}
p(b,c) + p(\bar{b},c)=1
\end{equation}
and multiply the left-hand term by unity, again using Ci:
\begin{equation}
\left[p(a,bc) + p(\bar{a},bc)\right]p(b,c) + p(\bar{b},c) =1
\end{equation}
and applying B2,
\begin{equation}
p(ab,c) + p(\bar{a}b,c) + p(\bar{b},c) =1
\end{equation}
which rearranges (again using Ci) to give
\begin{equation}
p(\bar{a}b,c) = -p(ab,c) + p(b,c)
\end{equation}
which we insert in Eq.~\ref{middle}, obtaining
\begin{equation}
p(b,c)p(a,c) \geq -p(ab,c) + p(b,c)
\end{equation}
which rearranges to give
\begin{equation}
p(b,c) - p(b,c)p(a,c) \leq p(ab,c)
\end{equation}
and, invoking B1,
\begin{equation}
p(b,c) - p(b,c)p(a,c) \leq p(a,c)
\end{equation}
which rearranges to give
\begin{equation}
p(b,c) \leq \frac{p(a,c)}{1-p(a,c)}.
\label{support}
\end{equation}

For $p(a,c) \geq 1/2$, that is an original event at least as probable as
its contradiction, this is no restriction on the supporting information
$b$ at all; which is what one would expect.  If $p(a,c)$ is small,
however, condition \ref{support} is very limiting.  
For very improbable original events
$p(b,c) \simeq p(a,c)$, which (recalling that $b$ must support $a$)
essentially makes $b$ identical with $a$.  Popper
 has not succeeded in
implementing his requirement of absolute support.

To be fair, there is no indication that he developed his formulation of
probability with this requirement in mind; he does not refer to it in
his exposition of the system.  We may say, however, that any success
he may have in remedying what he sees as the flaws in conventional
probability is less than total.

We now turn to the situations of zero-probability conditionals.
Unfortunately, we cannot yet apply Popper's system to either Haj\'{e}k's
problems or scientific hypotheses.  
It requires, as a conditional, $a\bar{a}$;
that is, a statement and its complement together, which not only have
probability zero of occuring simultaneously, but exhaust all possibilities
between them.  `A point is chosen at random on the surface of a sphere'
and `the point lands on the equator,' while having probability zero of
happening together, do not exhaust all possibilities.  `The point lands
on the equator' and `the point does not land on the equator' would
satisfy the requirement, but that's not the problem Haj\'{e}k sets.
Similarly, `an hypothesis is true' together with `the hypothesis is
false' would satisfy the condition, but again that's not what Popper
was aiming for.

(One might conceivably
postulate that {\em all} zero-probability statements in
Popper's system take the form of $\bar{a}a$.  It's not clear that it
could be done consistently, 
and in any case makes any interpretation as a system
of probability problematic.)

What Popper was clearly aiming for, and what Haj\'{e}k's problems and
`total functions' require, is a definite value for $p(a, bc)$ whenever
$p(b,c)=0$, and not only when $b=\bar{c}$. Popper desires a value
of unity; he asserts in \S 23, p. 91, p. 71,
`From a self-contradictory statement, any statement whatsoever can be
validly deduced.' (In the accompanying footnote he comments on others'
treatment of the idea.)  It is very similar to
Roeper and Leblanc's `total functions' region
(above), and thus is `widely adopted.'  Fitelson and Hawthorne [8]
agree that `everything follows from any inconsistent set of statements.'
The difference between contradictory statements and merely inconsistent
ones may seem trivial, but is important in applying Popper's 
mathematics. 

As Fitelson, Haj\'{e}k
and Hall [4] noted, Carnap ([10] pp. 295-6) considered the more
general prospect.  For statements conditional on logically false
(L-false) evidence, he notes that if they are not explicitly excluded, systems
such as Jeffreys' contain contradictions (though he allows that
Jeffreys may have tacitly excluded such statements).  The value
$P=1$ `seems most natural,' though some of Carnap's theorems 
must retain the exclusion of L-false conditionals even if this is
chosen.  If one assumes $P=0$ there are also problems, though different
ones\footnote{Everyone among the authories consulted has made the
tacit assumption that the probabilities of {\em all} statements
conditional on zero-probability events are identical.  This is
remarkable.  It does not hold for conditionals of any other value.
But further discussion would take us too far from the matter at
hand.}.  Carnap seems inclined to simply exclude L-false conditionals.

Given the clear intention of Popper and his supporters, let us
enumerate the possibilities.  The desired statement is
\vspace{0.5cm}

\noindent {\bf D}.  For any $c$ in $S$, and any $a$ and $b$ in $S$
such that $p(a,b)=0$, $p(c,ab)=1$.

\vspace{0.5cm}

Three things could happen:

\noindent (1) It could be that {\bf D} is inconsistent with 
Popper's formulation.

\noindent (2) It could be that {\bf D} is derivable in Popper's
system, and that the derivation simply doesn't appear in the
references at hand or has not yet been accomplished.

\noindent (3) It could be that {\bf D} is consistent with Popper's
formulation, but must be added as a separate axiom.

If (1) holds, Popper has accomplished the precise opposite of his
intention and what his supporters understand him to have done.  It is
also unclear why it shouldn't hold for $\bar{a}a$ conditionals.
We shall disregard this possibility.

If (2) or (3) holds the effect is the same.  Rather than spend what
might be a lot of time and effort attempting to determine which
is true, we will proceed to add
{\bf D} as an additional axiom, realizing that it might be unnecessary.

Applying it now to Haj\'{e}k's sphere, $c=$`the point lands in the
Western Hemisphere,', $a=$`the point lands on the Equator' and
$b=$`a point is chosen at random on a sphere.'  We find that the
probability of the point landing in the Western Hemisphere is 1,
which does not accord with Haj\'{e}k's intuition.  But we also
find, with probability 1, $c=$`the point lands in the Eastern
Hemisphere,' $c=$`the point lands on the Arctic Circle,' 
$c=$`the point does not land on the Equator.'  
{\em All} statements
are true; Haj\'{e}k's intuitive answers are there among them, but
so are their direct contradictions.  A similar situation holds
with his restricted series of coin flips.

(It might be possible to restrict the allowed $c$ to those Haj\'{e}k
desires, but to do so contradicts Popper's system as formulated,
and so is beyond the scope of this paper.)

Next we apply {\bf D} to a scientific theory, of (as Popper asserts)
zero probability, which was the original motivation
for Popper's formulation.  We find that {\em everything} is true, given
{\em any} theory.  All theories predict every event.  Thus, in an
enormous irony, Popper the exponent of falsification has produced
a system in which no theory can ever be falsified.

Of course Popper's derivations that the probabilities of all theories
vanish are erroneous.  That sends us back to the 
very beginning, with nothing in particular accomplished.

\section{Conclusions}

Popper's formulation of probability, motivated by his own errors in
applying the conventional theory, does not accomplish what he intended
and what his present supporters claim.
Like
a great deal of {\em The Logic of Scientific Discovery}, it has nothing
to do with the logic of scientific discovery.

\vspace{1.0cm}
\begin{center}
{\it References}
\end{center}

\noindent [1] Popper, Karl R. The Logic of Scientific Discovery,
(including Postscript: After Twenty Years). (New York: Harper \& Row),
1968 (original edition published 1934, in English 1958) \\

\noindent [2] Popper KR. The Logic of Scientific Discovery.
London, New York: Routledge, 2002. \\

\noindent [3] Whiting, AB, {\em Problems in the Science and
Mathematics of Popper's} The Logic of Scientific Discovery,
{\em Quanta}, volume 1, issue 1, pp. 13-18 (2012)
DOI: 10.12743/quanta.v1i1.3  \\

\noindent [4] Brian Fitelson, Alan Haj\'{e}k and Ned Hall, {\em Probability},
p. 4,
in {\em Philosophy of Science: An Encyclopedia}, eds. Jessica Pfeifer
and Sahotra Sarkar, Routledge Press, 2006 \\

\noindent [5] Van McGee, {\em Learning the Impossible}, p. 181, in
Ellery Eells and Brian Skyrms eds., {\em Probability and Conditionals:
Belief Revision and Rational Decision}, Cambridge, Cambridge University
Press, 1994\\

\noindent [6] P. Roeper and H. Leblanc, {\em Probability Theory and
Probability Logic}, Toronto: University of Toronto Press, 1999 \\

\noindent [7] Alan Haj\'{e}k, {\em Conditional Probability}, in
Prasanta Brandhopadhyay, Malcolm Forster eds., {\em Philosophy of
Statistics}, Elsevier, 2011 \\

\noindent [8] Brian Fitelson and Jim Hawthorne, {\em The Wason Task(s)
and the Paradox of Confirmation}, in {\em Philosophical Perspectives},
ed. J. Hawthorne and J. Turner, V. 24, pp. 207-241, December 2010;
p. 210 \\

\noindent [9] Harold Jeffreys, {\em Theory of Probability},
Oxford: Clarendon Press, 1948\\

\noindent [10] Rudolf Carnap, {\em Logical Foundations of 
Probability}, Chicago: University of Chicago Press, 1950
\end{document}